# Title: Global patterns of sex-biased migrations in humans

Authors: Chuan-Chao Wang[1], Li Jin[1, 2, 3], Hui Li[1,*]

**Affiliations:**

1. State Key Laboratory of Genetic Engineering and MOE Key Laboratory of Contemporary Anthropology, School of Life Sciences, Fudan University, Shanghai 200433, China
2. CAS-MPG Partner Institute for Computational Biology, SIBS, CAS, Shanghai, China
3. Institute of Health Sciences, China Medical City, Taizhou, Jiangsu, China

* Correspondence to: lihui.fudan@gmail.com

## Abstract

A series of studies have revealed the among-population components of genetic variation are higher for the paternal Y chromosome than for the maternal mitochondrial DNA (mtDNA), which indicates sex-biased migrations in human populations. However, this phenomenon might be also an ascertainment bias due to nonrandom sampling of SNPs. To eliminate the possible bias, we used the whole Y chromosome and mtDNA sequence data of 491 individuals from the 1000 Genomes Project Phase I to address the sex-biased migration dispute. We found that genetic differentiation between populations was higher for Y chromosome than for the mtDNA at global scales. The migration rate of female might be three times higher than that of male, assuming the effective population size is the same for male and female.

## Introduction

Sex-biased migration refers to a higher female migration rate in human populations (Seielstad et al., 1998). A series of studies have revealed higher $F_{ST}$ values and lower diversities for the SNP and STR data of male-specific region of Y chromosome (MSY) than mtDNA within or among world-wide populations, which indicates that Y chromosomes tend to be more localized geographically (Seielstad et al., 1998; Oota et al., 2001; Nasidze et al., 2004; Destro-Bisol et al., 2004; Wen et al., 2004; Wood et al., 2005; Li et al., 2013). Most interesting is Stoneking's observation (Oota et al., 2001) that the $F_{ST}$ value of mtDNA in matrilocal population of Northern Thailand is more than two times higher than that of MSY, thus, this discrepancy is probably caused by residence patterns. The majority of human societies are patrilocal, in which a married couple tends to reside with or near the husband's parents. As a result, women move more frequently than do men between populations or groups, leading to the among-population differences for the paternal Y chromosome is greater than that of the maternal mtDNA (Stoneking, 1998). The patrilocality might be reasonable to explain the sex-biased genetic difference on a local scale. But now comes the question, how can patrilocality shape the continental or even global-scale sex-biased genetic patterns observed by Seielstad et al. (Seielstad et al., 1998)? Actually, there has been a debate about whether the SNPs in the genealogical tree of Y chromosome could be used for diversity inference. SNPs are usually ascertained in small numbers of males and then genotyped in much larger population samples. This nonrandom sampling of SNPs can result in an ascertainment bias, which might overestimate the diversity in the populations in which those SNPs are discovered (Jobling, 2012). In 2004, Michael F. Hammer directly compared 6.7 kb of Y-chromosome Alu region and 770 bp of the mitochondrial CO3 gene in 389 individuals from ten globally diverse human populations to see whether the global-scale patterns of sex-biased migrations really exist. Hammer found that within-continent and between-continent genetic differentiations for Y chromosome and mtDNA were all similar, which suggest broader-scale genetic patterns might not always be caused by residence patterns (Wilder et al., 2004). However, Hammer's finding was probably biased as he only sequenced quite small region of Y chromosome and only a specific gene in mtDNA. Technological advances in sequencing now allow access to very large amounts of genetic data, which is eliminating this bias. For instance, the 1000 Genomes Project has sequenced the whole genomes for more than 1000 individuals. Here, we analyzed the whole Y chromosome and mtDNA sequence data of 491 individuals from the 1000 Genomes Project Phase I to address the sex-biased migration dispute.

## Materials and methods

The whole Y chromosome and mtDNA sequence data of 491 individuals from 13 worldwide populations were extracted from the 1000 Genomes Project Phase 1 (Table S1). The apportionment of diversity was calculated using the analysis of molecular variance (AMOVA), implemented in the Arlequin (version 3.5.1.3) (Excoffier & Lischer, 2010). To minimize biases associated with a higher mutation rate for mtDNA, we calculated genetic distances using a Tamura-Nei distance with high among-site rate heterogeneity, and for the Y chromosome, we calculated genetic distance using a Jukes-Cantor model of nucleotide substitution as suggested in (Wilder et al., 2004).

To investigate the migration rate of paternal Y chromosome and maternal mtDNA, we used the following method. Based on the island model of migration, $F_{ST}$ (among populations variation) is equal to $1/(1+Nv)$ for haploid systems (Y chromosome and mtDNA data), where N is the effective population size and v is equal to $m+\mu-m\mu$ (m: migration rate, μ: mutation rate) (Cavalli-Sforza & Bodmer, 1971; Seielstad et al., 1998).

## Results and Discussion

At global level, we observed the among-population component of genetic variation was more than two times higher for Y chromosome than for the mtDNA (Table 1). An AMOVA that used a hierarchical grouping of populations within continents gave the similar results. Values of $F_{ST}$ and $F_{CT}$ (among continents) were more than two times higher for Y chromosome than for the mtDNA, and $F_{SC}$ (among populations within continents) was even five times higher (Table 2). The interesting finding is the continental differences in the comparison of Y chromosome and mtDNA. The between-population component of genetic variation was slightly higher for mtDNA than for the Y chromosome in America and Africa (by 5~6 times), but not in Europe and Asia. In Europe and Asia, between-population component of genetic variation was about 10~20 times higher for Y chromosome than for the mtDNA (Table 1). Estimates of Nv from $F_{ST}$ are shown in Table S1. If the effective population size is the same for males and females, the female migration rate would exceed the male migration rate by a factor of 3.07.

We detect the signal of a higher migration rate among populations for females than for males at global scale. However, this conclusion is based on the assumption that the effective population sizes for male and female are equal. In fact, polygyny, higher rates of male mortality, or a greater variance in male lifetime reproductive success can decrease the effective population size of males (Wang et al., 2013). Whether the discrepancy in effective population size between male and female can account for the three times among-population genetic variation still needs further discussion.

## Acknowledgments

This work was supported by the National Excellent Youth Science Foundation of China (31222030), National Natural Science Foundation of China (31071098, 91131002), Shanghai Rising-Star Program (12QA1400300), Shanghai Commission of Education Research Innovation Key Project (11zz04), and Shanghai Professional Development Funding (2010001).

**Table 1** AMOVA results for the Y chromosome and mtDNA at population and continental scales

| Geographic region | Variation within population (%) | Variation among populations within groups (%) |
|---|---|---|
| **Global** | | |
| Y chromosome | 61.15 | 38.85 |
| mtDNA | 83.09 | 16.91 |
| **Europe (GBR, FIN, IBS, CEU, TSI)** | | |
| Y chromosome | 80.81 | 19.19 |
| mtDNA | 99.16 | 0.84 |
| **Asia (CHB, CHS, JPT)** | | |
| Y chromosome | 86.27 | 13.73 |
| mtDNA | 98.51 | 1.49 |
| **America (CLM, PUR, MXL)** | | |
| Y chromosome | 98.68 | 1.32 |
| mtDNA | 93.94 | 6.06 |
| **Africa (YRI, LWK)** | | |
| Y chromosome | 99.35 | 0.65 |
| mtDNA | 93.49 | 6.51 |

**Abbreviation:** GBR-British in England and Scotland; FIN-Finnish in Finland; IBS-Iberian populations in Spain; CEU-Utah residents with Northern and Western European ancestry; TSI-Toscani in Italy; CHB-Han Chinese in Bejing, China; CHS-Southern Han Chinese, China; JPT-Japanese in Tokyo, Japan; CLM-Colombian in Medellin, Colombia; PUR-Puerto Rican in Puerto Rico; MXL-Mexican Ancestry in Los Angeles, California; YRI-Yoruba in Ibadan, Nigeria; LWK-Luhya in Webuye, Kenya.

**Table 2** AMOVA grouping populations by continent of origin

| Locus | $F_{ST}$ (within populations %) | $F_{SC}$ (among populations within continents %) | $F_{CT}$ (among continents %) |
|---|---|---|---|
| Y chromosome | 0.43326 (56.67) | 0.11987 (7.72) | 0.35607 (35.61) |
| mtDNA | 0.19919 (80.08) | 0.02973 (2.45) | 0.17466 (17.47) |